\documentclass{PoS}

\usepackage{epsfig,amsmath}

\newcommand{\MZ}{M_{\rm Z}}
\newcommand{\GZ}{\Gamma_{\rm Z}}
\newcommand{\MH}{M_{\rm H}}
\newcommand{\mt}{m_{\rm t}}
\newcommand{\gev}{\,\, \mathrm{GeV}}

\title{Electroweak precision tests in the LHC era and beyond}

\ShortTitle{Electroweak precision tests}

\author{\speaker{Ayres Freitas}\\
        Pittsburgh Particle-physics Astro-physics \& Cosmology Center
(PITT-PACC), Department of Physics \& Astronomy, University of Pittsburgh,
Pittsburgh, PA 15260, USA\\
        E-mail: \email{afreitas@pitt.edu}}

\abstract{The current status of electroweak precision tests after the discovery
of the Higgs boson
is reviewed, both from a phenomenological and from a theoretical point of view.
Predictions for all $Z$-pole quantities are now
available at the complete fermionic two-loop order within the Standard Model.
The calculation of these corrections is described based on the example of the
total \mbox{$Z$-boson} width. Finally, an outlook on the experimental improvements and
theoretical challenges for a future high-luminosity $e^+e^-$ collider is given.}

\FullConference{ Loops and Legs in Quantum Field Theory - LL 2014,\\
		 27 April - 2 May 2014 \\
		 Weimar, Germany }

\begin{document}


\section{Introduction}

\begin{figure}[b]
\centering
\includegraphics[width=10cm, trim=0 0 0 400]{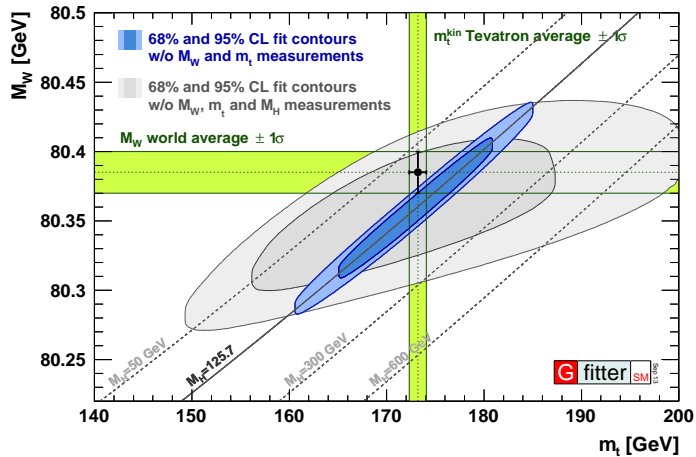}
\caption{Comparison of direct determinations (green) of $\mt$ and $M_{\rm
W}$ with the indirect determination from electroweak precision data before
(gray) and after (blue) the Higgs discovery. From Ref.~\cite{gfitter}.}
\label{fig:mtmw}
\end{figure}

With the recent discovery of the Higgs boson \cite{higgs} and first measurements
of its properties, the Standard Model (SM) has reached an unprecedented level
of experimental confirmation. All free parameters of the SM have been now
measured directly, and the agreement with indirect predictions from electroweak
precision observables is highly non-trivial. A global electroweak SM fit
including direct and indirect observables gives a high p-value of about 20\%, see
$e.\,g.$ Ref.~\cite{gfitter}.
For instance, the direct
measurements for the top-quark mass and $W$-boson mass are $\mt = 173.24
\pm 0.81$~GeV \cite{mte} and $M_{\rm W} = 80385 \pm 15$~MeV \cite{mwe},
respectively, while the fit to precision data yields $\mt = 177.0 \pm
2.1$~GeV and $M_{\rm W} = 80358 \pm 7$~MeV, see also Fig.~\ref{fig:mtmw}.

The electroweak precision fit uses inputs from a variety of sources, most
notably the measurement of the muon decay constant (which leads to the
strongest indirect constraint on $M_{\rm W}$) and properties of the $Z$-boson
from measurements of $e^+e^-$ collisions at $\sqrt{s} \approx \MZ$
(which are very sensitive to $\mt$ and the Higgs boson mass $\MH$).
These quantities have been measured with uncertainties of ${\cal O}(0.1\%)$ or
less. To match this precision on the theory side, one- and two-loop radiative
corrections, as well as dominant higher-order contributions must be included.

Due to the efforts of several groups over many years, complete two-loop
corrections are available for $M_{\rm W}$ \cite{mw} and the effective leptonic
weak mixing angle $\sin^2 \theta^\ell_{\rm eff}$ \cite{swl}, which is extracted
from the $Z$-pole left-right and forward-backward asymmetries. For the $Z$-boson
width ($\GZ$) and branching ratios, and the hadronic pole cross-section
$\sigma^0_{\rm had} \equiv \sigma[e^+e^- \to Z \to \text{hadrons}]$, the ${\cal
O}(\alpha\alpha_{\rm s})$ \cite{aas} and ${\cal O}(N_f\alpha^2)$ \cite{gz}
corrections have been computed, where $N_f$ stands for ``fermionic''
contributions from diagrams with at least one closed fermion loops, which are
expected to be dominant compared to the ``bosonic'' two-loop corrections. In
addition, universal leading 3- and 4-loop corrections for large values of $\mt$,
of ${\cal O}(\alpha_{\rm t}\alpha_{\rm s}^2)$, ${\cal O}(\alpha_{\rm
t}^2\alpha_{\rm s})$, ${\cal O}(\alpha_{\rm t}^3)$ and ${\cal O}(\alpha_{\rm
t}\alpha_{\rm s}^3)$, are known for all aforementioned observables \cite{3l}.
Here the abbreviation $\alpha_{\rm t} = \alpha \mt^2$ has been used. As evident
from Tab.~\ref{tab:prec}, the theory errors from missing higher-order
corrections are safely below current experimental uncertainties. 
\begin{table}[tb]
\centering
\begin{tabular}{lccccc}
\hline
& $M_{\rm W}$ & $\GZ$ & $\sigma^0_{\rm had}$ & $R_{\rm b}$ & 
 $\sin^2 \theta^\ell_{\rm eff}$\\
\hline
Exp.\ error & 15 MeV & 2.3 MeV & 37 pb & $6.6\times 10^{-4}$ & 
 $1.6\times 10^{-4}$ \\
Theory error  & 4 MeV  & 0.5 MeV & 6 pb & $1.5\times 10^{-4}$ &
 $0.5\times 10^{-4}$ \\
\hline
\end{tabular}
\caption{Current experimental errors and theory uncertainties for the SM
prediction of some of the most important electroweak precision observables. Here
$R_b \equiv \Gamma[Z \to b\bar{b}]/\Gamma[Z \to \text{hadrons}]$.}
\label{tab:prec}
\end{table}


\section{\textit{Z}-boson width at two loops}

As a concrete example for the electroweak two-loop corrections to electroweak
precision observables, this section will discuss the calculation of the ${\cal
O}(N_f\alpha^2)$ contribution to the (partial) $Z$-boson width(s). The total
$Z$-width is defined through the imaginary part of the complex pole of the
$Z$-boson propagator,
\begin{equation} s_0=\MZ^2-i\MZ\GZ. \label{eq:cpol} \end{equation}
This definition leads to a Breit-Wigner function with constant width near the
$Z$-pole, $\sigma \propto \linebreak |s-s_0|^{-2} = [(s-\MZ^2)^2+\MZ^2\GZ^2]^{-1}$. Note
that this differs from the Breit-Wigner function with a \emph{running} width
used in the experimental analyses, so that one has to include a finite shift
when relating $\MZ$ and $\GZ$ to the reported measured values:
\begin{equation}
\MZ = \MZ^{\rm exp} - 34.1 \text{ MeV}, \qquad
\GZ = \GZ^{\rm exp} - 0.9 \text{ MeV}.
\end{equation}
Expanding \eqref{eq:cpol} up to next-to-next-to-leading order (NNLO) and using
the power counting $\GZ \sim {\cal O}(\alpha)\MZ$, the result
for $\GZ$ can be written as \cite{gz}\footnote{Here a term $\propto
\text{Im}\,\Sigma''_{\rm Z}$ has been omitted, since $\text{Im}\,\Sigma''_{\rm
Z}=0$ at leading order for massless final-state fermions.}
\begin{equation}
\GZ = \frac{1}{\MZ}\, \text{Im}\,\Sigma_{\rm Z}(s_0)
 = \frac{1}{\MZ} \biggl [ \frac{\text{Im}\,\Sigma_{\rm Z}}{1+
  \text{Re}\,\Sigma'_{\rm Z}}\biggr ]_{s=\MZ^2} + {\cal O}(\GZ^3),
\end{equation}
where $\Sigma_{\rm Z}$ is the $Z$ self-energy. Using the optical theorem, the
imaginary part of the self-energy can be related to the decay process $Z \to
f\bar{f}$, resulting in
\begin{align}
\GZ = \sum_f \Gamma_f, \quad
\Gamma_f = \frac{N_c^f\MZ}{12\pi}\bigl [{\cal R}_V^f F_V^f +
{\cal R}_A^f F_A^f \bigr ]_{s=\MZ^2}, \quad
F_V^f \approx \frac{|v_f|^2}{1+\text{Re}\,\Sigma'_{\rm Z}},
\quad
F_A^f \approx \frac{|a_f|^2}{1+\text{Re}\,\Sigma'_{\rm Z}}, \label{eq:ff}
\end{align}
where $N_c^f=3(1)$ for quarks (leptons). Here the functions ${\cal
R}_{V,A}^f$ have been introduced, which capture effects from final-state QED and
QCD corrections. They are known up to ${\cal O}(\alpha_{\rm s}^4)$, ${\cal
O}(\alpha\alpha_{\rm s})$ and ${\cal O}(\alpha^2)$ in the limit of massless
fermions, while mass corrections are known up to three-loop order \cite{rad}.
The electroweak corrections are contained in $\Sigma'_{\rm Z}$ and the effective
$Zf\bar{f}$ vector and axial-vector couplings $v_f$ and $a_f$. 
Note that $v_f$ and $a_f$ include contributions from
photon-$Z$ mixing. Eq.~\eqref{eq:ff} is accurate up to NNLO.

For the calculation of the fermionic electroweak ${\cal O}(\alpha^2)$
corrections, Feynman diagrams have been generated with {\sl FeynArts 3.3}
\cite{feynarts}. In addition to the diagrams for the $Z \to f\bar{f}$ vertex
corrections, one also needs two-loop self-energy diagrams for the on-shell
renormalization \cite{mwlong}. In the on-shell renormalization scheme used here,
particle masses are defined through the (complex) pole of the propagators. For
the electroweak NNLO contributions, only $\MZ$, $M_{\rm W}$, $\MH$ and
$\mt$ are taken non-zero, while all other fermion masses have been
neglected (except for the final-state QED and QCD corrections, where also
non-vanishing bottom, charm and tau masses have been taken into account). The
electromagnetic coupling is renormalized in the Thomson limit, $i.\,e.$ for zero
momentum exchange, while $\alpha_{\rm s}$ is defined in the
$\overline{\text{MS}}$ scheme.

Two-loop integrals with a sub-loop self-energy bubble can be easily reduced to a
small set of master integrals using a generalization of the Passarino-Veltman
method \cite{pv} and integration-by-parts identities \cite{ibp}. These master
integrals can be evaluated numerically in terms of simple one-dimensional
integrations \cite{disp}
\begin{equation}
\raisebox{-10mm}{\includegraphics[width=3.5cm, viewport=210 450 330 530,
clip=true]{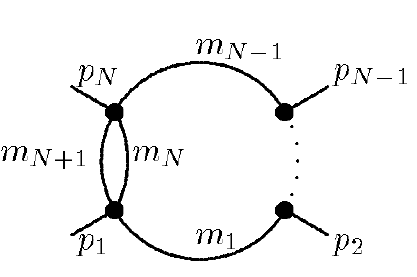}} = -2\int_{(m_N+m_{N+1})^2}^\infty ds \;
\text{Im}[B_0(s,m_N^2,m_{N+1}^2)] \, I^{(1)}(p_1,...,p_N;s,m_1^2,...,m_{N-1}^2),
\nonumber
\end{equation}
where $B_0$ is the usual one-loop two-point integral, while $I^{(1)}(\{p_i\};\{m_j\})$ is
a general scalar one-loop integral with momenta $\{p_i\}$ and masses $\{m_j\}$.

For two-loop vertex integrals with a triangle sub-loop, a technique based on
numerical integration over Feynman parameters has been used \cite{sub}. Before
the numerical integral can be carried out, potential divergences must be
subtracted. The global ultraviolet (UV) divergence of an integral $I^{(2)} =
\int d^dk_1d^dk_2 \; G(k_1,k_2,\{p_i\})$ is extracted by the operation
\begin{equation}
I^{(2)} = \int d^dk_1d^dk_2 \, \bigl [G(k_1,k_2,\{p_i\}) -
 G(k_1,k_2,\{0\})\bigr ]
 + \int d^dk_1d^dk_2 \; G(k_1,k_2,\{0\})
\equiv I^{(2)}_{\rm gs} + I^{(2)}_{\rm glob}. \label{eq:glob}
\end{equation}
Then $I^{(2)}_{\rm gs}$ is free of global UV singularities, while $I^{(2)}_{\rm
glob}$ is a two-loop vacuum integral, which can be solved analytically
\cite{vac}. $I^{(2)}_{\rm gs}$ may still contain sub-loop UV singularities.
After introducing Feynman parameters for the divergent sub-loop (assumed to be the $k_1$
loop), shifting the loop momentum, and appropriately canceling $k_1$ between
the numerators and denominator, the integral has the form
\begin{equation}
I^{(2)}_{\rm gs} = \int_0^1 dx_1 \dots dx_{m-1} \int d^dk_1 d^dk_2 \,
 \biggl [ \frac{C_1}{[k_1^2-A]^m} + \frac{C_2}{[k_1^2-A]^{m-1}} + \cdots +
 \frac{C_{m-1}}{[k_1^2-A]^2}\biggr ],
\end{equation}
where $A$ and the $C_i$ can depend on $k_2$ but not $k_1$. The sub-loop UV
divergence is contained in the last term $\propto C_{m-1}$ and can be removed
with the subtraction term $I^{(2)}_{\rm sub} = \int_0^1 dx_1 \dots dx_{m-1}
\int d^dk_1 d^dk_2 \; \frac{C_{m-1}}{[k_1^2-\mu^2]^2}$, where $\mu^2$ is a suitably chosen constant
parameter. Since $\mu^2$ is constant, the $k_1$ and Feynman parameter
integration of $I^{(2)}_{\rm sub}$ can be trivially performed analytically,
whereas $I^{(2)}_{\rm gs}-I^{(2)}_{\rm sub}$ is now UV-finite.

Infrared divergences can also be handled with suitable subtraction terms
\cite{sub}, but for the present calculation a photon mass has been used instead,
see Refs.~\cite{gz,swlept2} for details.

After also introducing Feynman parameters for the $k_2$ loop, the Feynman
parameter integration of the subtracted integral is performed numerically. In
the presence of physical thresholds, the integrand will in general contain
additional singularities where the denominator term $A$ vanishes. These internal singularities are
formally integrable but lead to difficulties for standard integration
algorithms. However, they can be avoided with a complex variable
transformation \cite{contour} 
\begin{equation}
x_i = z_i -i\lambda z_i(1-z_i) \frac{\partial A}{\partial x_i}\Big
|_{\vec{x}=\vec{z}}, \qquad 0 \leq z_i \leq 1,
\end{equation}
so that
\begin{equation}
A(\vec{x}) = A(\vec{z}) -i\lambda \sum_iz_i(1-z_i) \Bigl (\frac{\partial
A}{\partial x_i}\Bigr )^2_{\vec{x}=\vec{z}} + {\cal O}(\lambda^2).
\end{equation}
For sufficiently small $\lambda$, one can see that $A(\vec{x})$ will never
vanish as long as there is no point where $A(\vec{z}) = \nabla A(\vec{z}) = 0$
at the same time. More details of the calculation can be found in
Ref.~\cite{gz}.


\section{\textit{Z}-boson width: Results}

In the following, numerical results for the form factors $F^f_{V,A}$ from
\eqref{eq:ff} will be presented. For this purpose, the vector and
axial-vector components of the electroweak one-loop and fermionic two-loop
corrections were combined with the ${\cal O}(\alpha\alpha_{\rm s})$
contributions from Ref.~\cite{aas}, which also had to be split into vector and
axial-vector parts. Furthermore, the universal leading higher-order corrections
of ${\cal O}(\alpha_{\rm t}\alpha_{\rm s}^2)$, ${\cal
O}(\alpha_{\rm t}^2\alpha_{\rm s})$, ${\cal O}(\alpha_{\rm t}^3)$ and ${\cal
O}(\alpha_{\rm t}\alpha_{\rm s}^3)$ \cite{3l} have been included. 
The results are expressed in terms of the on-shell masses $\MZ$, $\MH$ and
$\mt$, the $\overline{\text{MS}}$ strong coupling constant $\alpha_{\rm
s}(\MZ)$, and the shift $\Delta\alpha$ of the electromagnetic coupling between
the scales $q^2=0$ and $\MZ^2$, $\Delta\alpha = 1-\alpha(0)/\alpha(\MZ^2)$. Note
that the $W$-mass is computed using the SM prediction for $M_{\rm
W}$ \cite{mw}.

The results can be conveniently expressed in terms of a simple parametrization
formula
\begin{align}
&F_X^f = \begin{aligned}[t] 
&F_0 + a_1 L_{\rm H} + a_2 L_{\rm H}^2 + a_3 \Delta_{\rm H} + a_4 \Delta_{\rm H}^2
 + a_5 \Delta_{\rm t} + a_6 \Delta_{\rm t}^2 + a_7 \Delta_{\rm t} L_{\rm H} \\
 &+ a_8 \Delta_{\alpha_{\rm s}} + a_9 \Delta_{\alpha_{\rm s}} L_{\rm H} + a_{10} \Delta_{\alpha_{\rm s}} \Delta_{\rm t} 
 + a_{11} \Delta_\alpha + a_{12} \Delta_{\rm Z}, 
 \label{eq:para} \end{aligned} \\[1ex]
&L_{\rm H} = \log\frac{\MH}{125.7\gev}, \quad
 \Delta_{\rm H} = \frac{\MH}{125.7\gev}-1,\quad
 \Delta_{\rm t} = \Bigl (\frac{\mt}{173.2\gev}\Bigr )^2-1, \quad
 \nonumber \\
& \Delta_{\alpha_{\rm s}} = \frac{\alpha_{\rm s}(\MZ)}{0.1184}-1, \quad
 \Delta_\alpha = \frac{\Delta\alpha}{0.059}-1, \quad
 \Delta_{\rm Z} = \frac{\MZ}{91.1876\gev}-1. \nonumber
\end{align}
Table~\ref{tab:res} shows the values of the coefficients obtained from a fit of
\eqref{eq:para} to the full result.
\begin{table}[p]
\renewcommand{\arraystretch}{1.2}
\small
\rule{0mm}{0mm}{\hspace{2.5em}}
\begin{tabular}{lrrrrrrr}
\hline
Form factor & \multicolumn{1}{c}{$F_0$} & 
 \multicolumn{1}{c}{$a_1$} & \multicolumn{1}{c}{$a_2$} & 
 \multicolumn{1}{c}{$a_3$} & \multicolumn{1}{c}{$a_4$} & 
 \multicolumn{1}{c}{$a_5$} & \multicolumn{1}{c}{$a_6$} \\
\hline
$F^\ell_V\;[10^{-5}]$ &
 19.84 & $-$1.012 & $-$0.1654 & 0.1467 & $-$0.00440 & 7.449 & 1.47 \\
$F^\ell_A\;[10^{-5}]$ &
 3446.44 & $-$3.2302 & $-$2.4615 & 2.1482 & $-$0.05666 & 25.856 & $-$2.59 \\
$F^\nu_{V,A}\;[10^{-5}]$ &
 3456.63 & $-$3.2844 & $-$2.4733 & 2.1561 & $-$0.05674 & 26.075 & $-$2.69 \\
$F^{u,c}_V\;[10^{-5}]$ &
 505.88 & $-$4.3145 & $-$1.013 & 0.9632 & $-$0.02295 & 23.31 & $-$1.79 \\
$F^{u,c}_A\;[10^{-5}]$ &
 3448.32 & $-$3.2438 & $-$2.4645 & 2.1502 & $-$0.05668 & 25.982 & $-$2.66 \\
$F^{d,s}_V\;[10^{-5}]$ &
 1650.61 & $-$5.0642 & $-$1.7806 & 1.6318 & $-$0.04038 & 29.384 & $-$3.15 \\
$F^{d,s}_A\;[10^{-5}]$ &
 3450.99 & $-$3.2528 & $-$2.4664 & 2.1514 & $-$0.05669 & 25.824 & $-$2.63 \\
$F^b_V\;[10^{-5}]$ &
 1620.96 & $-$4.7874 & $-$1.7196 & 1.5915 & $-$0.04005 & $-$3.582 & $-$1.97 \\
$F^b_A\;[10^{-5}]$ &
 3408.18 & $-$2.9546 & $-$2.4005 & 2.1073 & $-$0.05631 & $-$21.79 & $-$1.02 \\
\hline
\end{tabular}\\[1ex]
\rule{0mm}{0mm}{\hspace{2.5em}}
\begin{tabular}{lrrrrrrcc}
\hline
Form factor & \multicolumn{1}{c}{$a_7$} & \multicolumn{1}{c}{$a_8$} & 
 \multicolumn{1}{c}{$a_9$} & \multicolumn{1}{c}{$a_{10}$} & 
 \multicolumn{1}{c}{$a_{11}$} & \multicolumn{1}{c}{$a_{12}$} && max.\ dev. \\
\hline
$F^\ell_V\;[10^{-5}]$ &
 $-$0.278 & $-$0.13 & $-$0.17 & 2.6 & $-$42.7 & 1373 && $0.05$ \\
$F^\ell_A\;[10^{-5}]$ &
 0.0010 & $-$3.80 & $-$0.44 & 7.06 & $-$1.4 & 6915 && $0.07$ \\
$F^\nu_{V,A}\;[10^{-5}]$ &
 $-$0.0005 & $-$3.86 & $-$0.44 & 7.06 & $-$1.2 & 6943 && $0.07$ \\
$F^{u,c}_V\;[10^{-5}]$ &
 $-$0.067 & $-$3.88 & $-$0.37 & 6.07 & $-$145.2 & 5639 && $0.05$ \\
$F^{u,c}_A\;[10^{-5}]$ &
 0.0004 & $-$4.79 & $-$0.43 & 6.97 & $-$1.3 & 6922 && $0.07$ \\
$F^{d,s}_V\;[10^{-5}]$ &
 0.028 & $-$5.52 & $-$0.45 & 7.55 & $-$131.2 & 7491 && $0.07$ \\
$F^{d,s}_A\;[10^{-5}]$ &
 0.0036 & $-$4.96 & $-$0.43 & 6.97 & $-$1.1 & 6927 && $0.07$ \\
$F^b_V\;[10^{-5}]$ &
 0.334 & $-$1.28 & $-$0.50 & 11.9 & $-$130.6 & 7446 && $0.07$ \\
$F^b_A\;[10^{-5}]$ &
 0.340 & 0.96 & $-$0.49 & 13.0 & $-$1.7 & 6917 && $0.07$ \\
\hline
\end{tabular}
\caption{Coefficients for the parametrization formula (3.1) for the
$Zf\bar{f}$ form factors. Within the ranges $70\gev < \MH < 1000\gev$, $165\gev < \mt < 190\gev$,
$\alpha_{\rm s}=0.1184\pm 0.0050$, $\Delta\alpha = 0.0590 \pm 0.0005$ and $\MZ = 91.1876 \pm
0.0084 \gev$, the formula approximates the full result with maximal deviations
given in the last column. When restricting oneself to the range $\MH = 125.7\pm
2.5\gev$, the maximum deviation is reduced by a factor of more than 5.}
\label{tab:res}
\end{table}

Results for the complete observables, $i.\,e.$ the total and partial widths
including the final state radiation functions ${\cal R}_{V,A}^f$, have been
presented elsewhere \cite{gz}.


\section{Outlook}

Electroweak precision observables are a very useful for accurate indirect tests
of the SM and setting stringent constraints on new physics. At the current level
of experimental precision, they are sensitive to one- and two-loop and even
leading higher-order effects. Nevertheless, due to tremendous efforts by many
groups, the error of the theoretical predictions within the SM has been reduced
comfortably below the experimental uncertainty for the most relevant quantities,
see Tab.~\ref{tab:prec}. The LHC experiments are expected to provide interesting
independent determinations of $M_{\rm W}$ and $\sin^2 \theta^\ell_{\rm eff}$
\cite{lhc}, but the overall experimental precision for these quantities will not
improve markedly.

However, a future high-luminosity $e^+e^-$ machine like ILC will substantially
increase the experimental precision, thus posing a challenge for theorists to
match this precision. In Table~\ref{tab:ilc}, the expected ILC precision is compared
with the current and projected theory uncertainty for several observables. For
the projection, it is assumed that the leading fermionic three-loop corrections
of order ${\cal O}(N_f^2\alpha\alpha_{\rm s})$, ${\cal O}(N_f\alpha\alpha_{\rm
s})$, ${\cal O}(N_f^3\alpha^3)$ and ${\cal O}(N_f^2\alpha^3)$ will be computed,
where $N_f$ indicates the number of closed fermion loops. These contributions
imply three-loop self-energies and three-loop vertices with sub-loop bubbles.
The remaining theory error is estimated by approximating the perturbation
series with a geometric series.
\begin{table}[tb]
\centering
\begin{tabular}{lcccc}
\hline
& $M_{\rm W}$ & $\GZ$ & $R_{\rm b}$ & 
 $\sin^2 \theta^\ell_{\rm eff}$\\
\hline
ILC exp.\ error & 3\dots 5 MeV & $\sim$1 MeV & $1.5\times 10^{-4}$ & 
 $1.3\times 10^{-5}$ \\
Current theory error  & 4 MeV  & 0.5 MeV & $1.5\times 10^{-4}$ &
 $4.5\times 10^{-5}$ \\
Projected theory error  & 1 MeV  & 0.2 MeV & $0.5\dots 1\times 10^{-4}$ &
 $1.5\times 10^{-5}$ \\
Parametric error for ILC & 2.6 MeV & 0.5 MeV & $< 10^{-5}$ & $2\times 10^{-5}$\\
\hline
\end{tabular}
\caption{Projected experimental errors of ILC running at $\sqrt{s} \approx \MZ$
and $\sqrt{s} \approx 2M_{\rm W}$ \cite{ilc} and current and expected future
theory uncertainties for the SM prediction for several important electroweak
precision observables. The future theory errors are estimated under the
assumption that ${\cal O}(N_f^2\alpha\alpha_{\rm s})$, ${\cal O}(N_f\alpha\alpha_{\rm
s})$, ${\cal O}(N_f^3\alpha^3)$ and ${\cal O}(N_f^2\alpha^3)$ corrections will become available. The
parametric error describes the uncertainty of the SM prediction due to
uncertainties of input parameters: $\delta \mt = 100$~MeV,
$\delta\alpha_{\rm s}=0.001$ (from ILC), and $\delta\MZ = 2.1$~MeV (from LEP).}
\label{tab:ilc}
\end{table}

As evident from the table, it seems not preposterous to believe that the theory
calculations can achieve a level of precision comparable or better than the
expected ILC precision. In fact, for most quantities, the parametric error due
to the uncertainty of input parameters will dominate over the perturbative
theory error. 

\paragraph{Acknowledgements:} 
This work has been supported in part by the National Science Foundation under
grant no.\ PHY-1212635.



\begin{thebibliography}{99}

\bibitem{higgs} 
  G.~Aad {\it et al.}  [ATLAS Collaboration],
  Phys.\ Lett.\ B {\bf 716} (2012) 1
  [arXiv:1207.7214 [hep-ex]];
  S.~Chatrchyan {\it et al.}  [CMS Collaboration],
  Phys.\ Lett.\ B {\bf 716} (2012) 30
  [arXiv:1207.7235 [hep-ex]].

\bibitem{gfitter}
  M.~Baak {\it et al.},
  Eur.\ Phys.\ J.\ C {\bf 72} (2012) 2205
  [arXiv:1209.2716 [hep-ph]],
  and updates at {\tt gfitter.desy.de}.

\bibitem{mte}
  M.~Muether {\it et al.}  [Tevatron Electroweak Working Group and CDF and D\O\  Collaborations],
  arXiv:1305.3929 [hep-ex];
  The ATLAS and CMS Collaborations,
  ATLAS-CONF-2013-102.

\bibitem{mwe}
  S.~Schael {\it et al.}  [ALEPH and DELPHI and L3 and OPAL and LEP Electroweak Collaborations],
  Phys.\ Rept.\  {\bf 532} (2013) 119
  [arXiv:1302.3415 [hep-ex]];
  T.~A.~Aaltonen {\it et al.}  [CDF and D\O\ Collaborations],
  Phys.\ Rev.\ D {\bf 88} (2013) 5,  052018
  [arXiv:1307.7627 [hep-ex]].

\bibitem{mw}
A.~Freitas, W.~Hollik, W.~Walter and G.~Weiglein,
Phys.\ Lett.\ B {\bf 495} (2000) 338
[Erratum-ibid.\ B {\bf 570} (2003) 260]
[hep-ph/0007091];
M.~Awramik and M.~Czakon,
Phys.\ Rev.\ Lett.\  {\bf 89} (2002) 241801
[hep-ph/0208113],
Phys.\ Lett.\ B {\bf 568} (2003) 48 
[hep-ph/0305248];
A.~Onishchenko and O.~Veretin,
Phys.\ Lett.\ B {\bf 551} (2003) 111
[hep-ph/0209010];
  M.~Awramik, M.~Czakon, A.~Freitas and G.~Weiglein,
  Phys.\ Rev.\ D {\bf 69} (2004) 053006
[hep-ph/0311148].

\bibitem{swl}
  M.~Awramik, M.~Czakon, A.~Freitas, G.~Weiglein,
  Phys.\ Rev.\ Lett.\  {\bf 93} (2004) 201805
    [hep-ph/0407317];
  M.~Awramik, M.~Czakon and A.~Freitas,
  Phys.\ Lett.\  B {\bf 642} (2006) 563
  [hep-ph/0605339];
  W.~Hollik, U.~Meier and S.~Uccirati,
  Nucl.\ Phys.\ B {\bf 731} (2005) 213
  [hep-ph/0507158],
  Nucl.\ Phys.\ B {\bf 765} (2007) 154
  [hep-ph/0610312].

\bibitem{aas}
  A.~Czarnecki and J.~H.~K\"uhn,
  Phys.\ Rev.\ Lett.\  {\bf 77} (1996) 3955
  [hep-ph/9608366];
  R.~Harlander, T.~Seidensticker and M.~Steinhauser,
  Phys.\ Lett.\ B {\bf 426} (1998) 125
  [hep-ph/9712228].

\bibitem{gz}
  A.~Freitas,
  Phys.\ Lett.\ B {\bf 730} (2014) 50
  [arXiv:1310.2256 [hep-ph]],
  JHEP {\bf 1404} (2014) 070
  [arXiv:1401.2447 [hep-ph]].

\bibitem{3l}
L.~Avdeev, J.~Fleischer, S.~Mikhailov and O.~Tarasov,
Phys.\ Lett.\ B {\bf 336} (1994) 560
[Erratum-ibid.\ B {\bf 349} (1994) 597] 
[hep-ph/9406363]; 
K.~G.~Chetyrkin, J.~H.~K\"uhn and M.~Steinhauser,
Phys.\ Lett.\ B {\bf 351} (1995) 331
[hep-ph/9502291];
J.~J.~van der Bij, K.~G.~Chetyrkin, M.~Faisst, G.~Jikia and T.~Seidensticker,
Phys.\ Lett.\ B {\bf 498} (2001) 156 
[hep-ph/0011373];
M.~Faisst, J.~H.~K\"uhn, T.~Seidensticker and O.~Veretin,
Nucl.\ Phys.\ B {\bf 665} (2003) 649
[hep-ph/0302275];
Y.~Schr\"oder and M.~Steinhauser,
Phys.\ Lett.\ B {\bf 622} (2005) 124
[hep-ph/0504055];
K.~G.~Chetyrkin, M.~Faisst, J.~H.~K\"uhn, P.~Maierhoefer and C.~Sturm,
  Phys.\ Rev.\ Lett.\  {\bf 97} (2006) 102003
  [hep-ph/0605201];
  R.~Boughezal and M.~Czakon,
  Nucl.\ Phys.\  B {\bf 755} (2006) 221
  [hep-ph/0606232].

\bibitem{rad}
  K.~G.~Chetyrkin, J.~H.~K\"uhn and A.~Kwiatkowski,
  Phys.\ Rept.\  {\bf 277}, 189 (1996);
P.~A.~Baikov, K.~G.~Chetyrkin and J.~H.~K\"uhn,
  Phys.\ Rev.\ Lett.\  {\bf 101}, 012002 (2008)
  [arXiv:0801.1821 [hep-ph]];
P.~A.~Baikov, K.~G.~Chetyrkin, J.~H.~K\"uhn and J.~Rittinger,
  Phys.\ Rev.\ Lett.\  {\bf 108}, 222003 (2012)
  [arXiv:1201.5804 [hep-ph]];
  A.~L.~Kataev,
  Phys.\ Lett.\ B {\bf 287}, 209 (1992).

\bibitem{feynarts}
  T.~Hahn,
  Comput.\ Phys.\ Commun.\  {\bf 140}, 418 (2001)
  [hep-ph/0012260].

\bibitem{mwlong}
A.~Freitas, W.~Hollik, W.~Walter and G.~Weiglein,
Nucl.\ Phys.\ B {\bf 632} (2002) 189
[Erratum-ibid.\ B {\bf 666} (2003) 305]
  [hep-ph/0202131].

\bibitem{pv}
  G.~Weiglein, R.~Scharf and M.~B\"ohm,
  Nucl.\ Phys.\ B {\bf 416}, 606 (1994)
  [hep-ph/9310358].

\bibitem{ibp}
K.~G.~Chetyrkin and F.~V.~Tkachov,
Nucl.\ Phys.\ B {\bf 192}, 159 (1981); 

\bibitem{disp}
S.~Bauberger, F.~A.~Berends, M.~B\"ohm and M.~Buza,
Nucl.\ Phys.\ B {\bf 434}, 383 (1995); 
[hep-ph/9409388];
S.~Bauberger and M.~B\"ohm,
Nucl.\ Phys.\ B {\bf 445}, 25 (1995)
[hep-ph/9501201].

\bibitem{sub}
  A.~Freitas,
  JHEP {\bf 1207}, 132 (2012)
  [Erratum-ibid.\  {\bf 1209}, 129 (2012)]
  [arXiv:1205.3515 [hep-ph]].

\bibitem{vac}
  A.~I.~Davydychev and J.~B.~Tausk,
  Nucl.\ Phys.\ B {\bf 397} (1993) 123.

\bibitem{swlept2}
  M.~Awramik, M.~Czakon and A.~Freitas,
  JHEP {\bf 0611}, 048 (2006)
   [arXiv:hep-ph/0608099].

\bibitem{contour}
  Z.~Nagy and D.~E.~Soper,
  Phys.\ Rev.\  D {\bf 74}, 093006 (2006)
  [hep-ph/0610028].

\bibitem{lhc}
  G.~L.~Bayatian {\it et al.}  [CMS Collaboration],
  J.\ Phys.\ G {\bf 34} (2007) 995;
  G.~Aad {\it et al.}  [ATLAS Collaboration],
  arXiv:0901.0512 [hep-ex].

\bibitem{ilc}
  R.~Hawkings and K.~M\"onig,
  Eur.\ Phys.\ J.\ direct C {\bf 1} (1999) 8
  [hep-ex/9910022];
  M.~Baak {\it et al.},
  arXiv:1310.6708 [hep-ph].

\end{thebibliography}
\end{document}